\documentclass[preprint,3p]{elsarticle}
\usepackage{epstopdf}
\usepackage{caption}
\usepackage{setspace}
\usepackage{amssymb}
\usepackage{multirow}
\usepackage{amsthm}
\usepackage{booktabs}
\usepackage{amsmath}
\usepackage{amssymb}
\usepackage{mathrsfs}
\usepackage{longtable}
\usepackage{lscape}
\usepackage{url}
\usepackage{tabularx}
\usepackage{epsfig}
\usepackage{colortbl}
\usepackage{subfigure}
\usepackage{tikz,mathpazo}
\usepackage{graphicx}
\usepackage{natbib}
\usepackage[noend]{algpseudocode}
\usepackage{algorithm}
\usepackage{algorithmicx}
\usepackage{hyperref}
\usepackage [latin1]{inputenc}

\newtheorem{definition}{Definition}[section]
\usetikzlibrary{shapes.geometric, arrows}
\biboptions{numbers,sort&compress}
\journal{}

\begin{document}
\begin{frontmatter}
\title{A modified gravity model based on network efficiency for vital nodes identification in complex networks}

\author[address1]{Hanwen Li}
\author[address1,address5]{Qiuyan Shang}
\author[address1,address2,address3,address4]{Yong Deng \corref{label1}}

\address[address1]{Institute of Fundamental and Frontier Science, University of Electronic Science and Technology of China, Chengdu 610054, China}
\address[address5]{Yingcai Honors of school, University of Electronic Science and Technology of China, Chengdu, 610054, China}
\address[address2]{School of Education, Shaanxi Normal University, Xi'an, 710062, China}
\address[address3]{School of Knowledge Science, Japan Advanced Institute of Science and Technology, Nomi, Ishikawa 923-1211, Japan}
\address[address4]{Department of Management, Technology, and Economics, ETH Zrich, Zurich, Switzerland}

\cortext[label1]{Corresponding author at: Institute of Fundamental and Frontier Science, University of
	Electronic Science and Technology of China, Chengdu, 610054, China. E-mail: prof.deng@hotmail.com, dengentropy@uestc.edu.cn.(Yong Deng)}
\begin{abstract}
Vital nodes identification is an essential problem in network science. Various methods have been proposed to solve this problem. In particular, based on the gravity model, a series of improved gravity models are proposed to find vital nodes better in complex networks. However, they still have the room to be improved. In this paper, a novel and improved gravity model, which is named network efficiency gravity centrality model (NEG), integrates gravity model and network efficiency is proposed.  Compared to other methods based on different gravity models, the proposed method considers the effect of the nodes on structure robustness of the network better. To solidate the superiority of the proposed method, experiments on varieties of real-world networks are carried out.
\end{abstract}
\end{frontmatter}
\section{Introduction}

Network science is an emerging research direction as it is an effective theory and tool which can be used in many real-environments like economy \cite{zhu2021analysis,gao2019computational,altinoglu2021origins}, biology \cite{liu2020computational,derakhshani2020gene,nomi2021complex}, transportation \cite{wang2019risk,gu2020performance,zhou2021resilience}, chemistry \cite{chrayteh2021disentangling}, social system \cite{wu2021maximum,tang2021delegation,zhou2020particle} and so forth. More and more researchers are devoted to network science \cite{le2020mining,dong2018consensus}. There are many worthy research topics in network science including but not limiting to community detection \cite{vega2020influence}, fractal dimension \cite{rak2020fractional,wen2021invited}, link prediction \cite{lee2021collaborative}, evolutionary game theory \cite{shen2021exit,chu2020evolution,wen2021alternating}, self similarity analysis \cite{wang2021self} and so forth. Algorithms and tools in network science can also be used for time series analysis \cite{huang2021natural,liu2020fuzzy}, pattern recognition \cite{mohammadpoory2019automatic,kong2021eeg}, multi-criteria decision making \cite{xiong2021conflicting,CHEN2021104438}, uncertainty modeling \cite{Zhao2020complex}, recommender system \cite{bai2020customer,tian2020novel}, just to name a few. We will see the emerging progress in both the theory and the applications of network science in the near future \cite{li2021deep,diocsan2021network,yu2020identifying,seong2022forecasting}.

Vital nodes identification is an important issue in network science as it can be used in many fields and directions like link prediction \cite{wu2019general}, disease control \cite{doostmohammadian2020centrality,luo2020new}, risk engineering \cite{wang2021robustness,umunnakwe2021cyber}, biology engineering \cite{grubb2021network,meng2021protein,lei2019random} and decision making \cite{li2021research,trach2021centrality}. Therefore, it is essential to study on vital nodes identification. There are various methods proposed for vital nodes identification \cite{qiu2021ranking}. Newman \cite{newman2005measure} considers the path information and propose betweenness centrality. Zhang et al. \cite{zhang2021lfic}  propose local fuzzy information centrality to identify vital nodes. Berahmand et al. \cite{berahmand2018new} investigate both of the positive and negative effect of local clustering coefficient for this problem. Maji et al. \cite{maji2021identifying} take the influence overlap into account. Inspired by random walk algorithm for link prediction \cite{liu2010link}, Ren et al. \cite{ren2014iterative} propose a new method called iterative resource allocation. Graph energy theory can also be used in vital nodes identification as Ma et al. \cite{ma2019quasi} propose quasi-laplacian centrality. Li $et$ $al.$ \cite{li2021generalized} propose a generalized gravity model considering the local information of the node more comprehensively. Liu et al. \cite{liu2020gmm} propose a generalized mechanism model, which is also called gravity model to identify vital nodes. Zhao et al. \cite{zhao2020theidentification} propose a novel method uses structure similarity. Wen et al.\cite{wen2020Vital} explores fractal theory and propose local multi-dimension to investigate vital nodes. Discrete Moth-Flame optimization algorithm is also utilized for solving this problem by Wang et al. \cite{wang2021identifying}. Gupta and Mishra \cite{gupta2021spreading} consider the network structure in this problem. Convolutional network is also proved to be efficient to find vital nodes by Yu et al. \cite{yu2020identifying}. A novel attraction measure is proposed by Wang et al. \cite{wang2021novel}. Hierarchical approach for influential node ranking is proposed by Zareie and Sheikhahmadi \cite{zareie2018hierarchical}. Greedy alrotihm \cite{zhang2020influential} and spinning trees \cite{dai2019influential} are also useful alternatives to investigate important nodes. What is more, some open source node centrality platform is developed \cite{zhuang2020complex}. More kinds of superior methods will be explored in the upcoming days. 

However, many methods still have room to be improved. In particular, since gravity models are very effective in identifying vital nodes in complex networks because gravity model can simulate the interaction of nodes and generalize both local and global information. Many modified gravity models have been proposed for further study. Whereas, most existing gravity models do not consider the impact of key nodes on the overall structure and robustness of the network. Therefore, an improved gravity model for vital nodes identification, which is called network efficiency gravity model is proposed in this paper. Compared to the previous modified gravity model, the proposed model is related to the network efficiency of the nodes and it considers the effect of the node on the whole network, which is more comprehensive. The superiority of the proposed method will be proven by experiments on many real-world networks.

The contribution of our method is as follows. First, the degree of a node contains the local neighborhood information of the node, and the path between nodes represents the global information. In the proposed model, the local information and the global information are comprehensively considered. Second, the gravity model regards the identification of important nodes as a global issue, and comprehensively considers the mutual influence between various nodes in the network. Third, the influence of nodes on network efficiency reflects the influence of nodes on the overall structure and robustness of the network. Fourth, the comprehensive use of the original gravity model and network efficiency-related indicators reflects the importance of nodes from the perspective of the mutual influence between nodes and the overall structure of the network.

The remainder structure of this paper is given as follows. The preliminaries are briefly introduced in Section 2. In Section 3, the modified gravity model is proposed. Experiments on real-world networks are utilized to show the superiority of the proposed method in the next section. In the end, the conclusion is given in Section 5.

\section{Preliminaries}

Some preliminaries including containing complex network, susceptible-infected model, Kendall correlation coefficient, network efficiency and classic identification approaches are briefly introduced as follows. 

\subsection{Complex network}
A complex network can be defined as a graph $G(V, E)$, where $V$ is the set of the vertices and $E$ is the set of the edges. $A$ is the adjacency matrix and $a_{ij}$ represents the edge between node $i$ and node $j$. If the node $i$ and node $j$ is connected, then $a_{ij} = 1$ otherwise 0. $d_{ij}$ refers to the shortest distance between node $i$ and $j$, and it can be calculated by shortest path algorithms. 

\subsection{Susceptible-infected model}

A representative and effective model that can simulate the process of disease propagation is the susceptible-infected (SI) model \cite{zhou2006behaviors}. Every node in this model has two states. They are either in susceptible or infected state. Every node of the network is susceptible in the beginning. Some nodes can be chosen to be infected in the model and then infect other nodes. $T$ refers to the simulation steps of the model. $K$ means the times of independent experiments. $N(t)$ means the average number of infected nodes in $K$ times of independent experiments. $\beta$ is the infection rate, which is the probability of the susceptible node is infected by other infected nodes.

\subsection{Kendall correlation coefficient}
Kendall correlation coefficient is a classical mathematical measure which can be used to measure the correlation of two variables \cite{kendall1945treatment}. It is defined as follows.

\begin{definition}
	Given two arrays $X = (X_1, X_2, ..., X_n)$, $Y = (Y_1, Y_2, ..., Y_n)$, the pairs $(X_i, X_j)$ and $(Y_i, Y_j)$ are called consistent if $X_i > X_j$ and $Y_i > Y_j$ otherwise they are inconsistent. The Kendall correlation coefficient can be calculated as follows.
	
	\begin{equation}
	tau (X,Y) = \frac{2(c^+ - c^-)}{n(n-1)}
	\end{equation}
	where $c^+$ means the number of consistent pairs and $c^-$ means the number of inconsistent pairs. 
\end{definition}

\subsection{Network efficiency}
Network efficiency is often utilized to describe the robustness of the network \cite{liu2020gmm}. It is defined as follows.

\begin{equation}
	E(G) = \frac{\sum\limits_{i \not= j}\frac{1}{d_{ij}}}{N (N-1)}
\end{equation}

where $N$ represents the number of nodes in the network and the length of the shortest path between node $i$ and node $j$ is denoted as $d_{ij}$.
\subsection{Centrality measures}
The baseline centrality measures used in this paper are briefly introduced as follows.

\subsubsection{Gravity model \cite{li2019identifying}}
Based on the idea of attraction between different nodes, a new method is proposed by Li et al. \cite{li2019identifying} for vital nodes identification, inspired by the gravitation law in physics \cite{rosenfeld1965newton}, which considers both global and local information of the nodes. It is calculated as follows.

\begin{equation}
	C_G(i) = \sum\limits_{i \not= j,d_{ij} \leq r}\frac{k_i \times k_j}{d_{ij}^2}
\end{equation}
where $k_i$ and $k_j$ refers to the degree of node $i$ and node $j$ respectively. $r$ is the truncation radius of the network and $r = \frac{\left\langle d\right\rangle}{2}$. The $\left\langle d\right\rangle$ is the average length of the
shortest paths of the network

\subsubsection{Weighted gravity model \cite{liu2020gmm}}
Liu et al. \cite{liu2020gmm} modified the gravity model and proposed the weighted gravity model, which also considers the effect of the node on other nodes. It is calculated as follows.

\begin{equation}
\begin{split}
	AX &= \lambda X\\
	e_i &= X_i\\
C_{WF} &= \sum\limits_{i \not= j,d_{ij} \leq r}e_i\frac{k_i \times k_j}{d_{ij}^2}
\end{split}
\end{equation}

where $X$ and $\lambda$ are normalized eigenvector and largest eigen vector respectively.

\subsubsection{Generalized gravity model \cite{li2021generalized}}
Generalized gravity model for identifying the vital nodes is proposed by Li et al. \cite{li2021generalized} as a generalization of gravity model, which is more flexible. It is calculated as follows.

\begin{equation}
C_{GG} = \sum\limits_{i \not= j, d_{ij} \leq r}e^{-\alpha LCC(i)} \times e^{-\alpha LCC(j)} \times \frac{k_i \times k_j}{d_{ij}^2}
\end{equation}
where $LCC(i)$ and $LCC(j)$ are the local clustering coefficient of the node $i$ and node $j$ respectively. Take node $i$ as an example, the local clustering coefficient of node $i$ can be calculated as

\begin{equation}
	LCC(i) = \frac{n_i}{k_i \times (k_i - 1)}
\end{equation}
where $n_i$ is the number of edges between the neighbors of the node $i$.

\subsubsection{Betweenness centrality \cite{newman2005measure}}
Betweenness centrality considers the path information of the node \cite{newman2005measure}, it is a global information based index. It is given as follows.

\begin{equation}
	C_{BC}(i)=\sum\limits_{j,k \not= i}\frac{P_{jk}(i)}{P_{jk}}
\end{equation}

where $P_{jk}$ refers to the number of shortest paths between node $j$ and $k$ and $P_{jk}(i)$ means the number of shortest paths between node $j$ and $k$ which also pass through node $i$.
\subsubsection{Iterative resource allocation \cite{ren2014iterative}}

Iterative resource allocation is a method based on random walk for vital nodes identification. Assuming resource is allocated the same at all nodes and then resources are transported in the network, the most important node should have the most resource after certain steps. This model can be calculated as follows. 

\begin{equation}
I(t+1)=XI(t)=\begin{pmatrix}
x_{11} & x_{12} & ... & x_{1n}\\
x_{21} & x_{22} & ... & x_{2n}\\
... & ... & ... & ...\\
x_{n1} & x_{n2} & ... & x_{nn}
\end{pmatrix}
\begin{pmatrix}{}
I_1(t)\\I_2(t)\\...\\I_n(t)
\end{pmatrix}
\end{equation}
where the $X$ means the transition matrix. $x_{ij}$ can be calculated as follows.

\begin{equation}
x_{ij} = a_{ij} \frac{C(j)}{\sum\limits_{i \not= j} C(j)}
\end{equation}
Where $a_{ij}$ is the corresponding element in the adjacent matrix of the edge between node $i$ and node $j$. $C(j)$ is the centrality of the node $j$, any centrality can be chosen as the input. For easy calculation, degree centrality \cite{newman2003structure} is selected in this paper beacuse degree centrality is derived from local information and it is easier to be obtained.
At the initial process, $I_1(1) = I_2(1) = ... = I_n(1) = 1$.

\subsubsection{Quasi-laplacian centrality \cite{gao2019computational}}

Quasi-laplacian centrality is proposed based on graph energy \cite{gao2019computational}, which can fully reveals the influence of a node on the whole network. And can be calculated as follows.

\begin{equation}
	\begin{split}
	E_{ql}(G)&=\sum\limits_{i} k_i +\sum\limits_{i} k_i^2\\
	C_{ql}(i) &= E_{ql}(G) - E_{ql}(G-i)
	\end{split}
\end{equation}
where $E_{ql}(G)$ refers to the graph quasi-laplacian energy of the network and $E_{ql}(G-i)$ refers to the graph quasi-laplacian energy of the network deletes node $i$. 

\section{Proposed approach}
In this section, a new method which is called network efficiency gravity model (NEG) is proposed. The proposed method is defined as follows.

\begin{equation}
C_{NEG}(i) = \frac{E(G)}{E(G-i)}\sum\limits_{i \not= j}\frac{k_i \times k_j}{d_{ij}^2}
\end{equation}
where $E(G-i)$ is the network efficiency of the network which deletes node $i$.

In the proposed model, From the perspective of the influence of nodes on the whole structure of the network, the more important the node $i$, the greater the influence on the network efficiency of the entire network after it is removed. That is to say, the more important the node $i$, the smaller $E(G-i)$ and the larger $\frac{E(G)}{E(G-i)}$. Consequently, a more vital node is of a larger $C_{NEG}(i)$. Compared to the most other gravity models, the proposed method considers the effect of the node on the whole network structure, which is more comprehensive and effective.

\section{Experiments}
In this section, experiments are carried out to demonstrate the superiority of the proposed method. The proposed method is compared with some effective methods including gravity model (G) \cite{li2019identifying}, weighted gravity model (WG) \cite{liu2020gmm}, generalized gravity model (GG) \cite{li2021generalized}, iterative resource allocation (IRA) \cite{ren2014iterative}, quasi-laplacian centrality (QL) \cite{ma2019quasi} and betweenness centrality \cite{newman2005measure}. Networks derived from the real-world are used to test whether or not the proposed method works. The networks used in this paper are Jazz (198 nodes, 2742 edges), NS (379 nodes, 914 edges), USAir (332 nodes, 2126 edges), PB (1222 nodes, 16714 edges), Infectious (410 nodes, 17298 edges), PDZbase (212 nodes, 2672 edges), SCC (151 nodes, 9800 edges) and Haggle274 (274 nodes, 28244 edges). The results are a little different from \cite{li2021generalized} due to code problems but do not affect the effectiveness of generalized gravity model.

\begin{figure}
	\centering
	\subfigure{\includegraphics[scale=0.45]{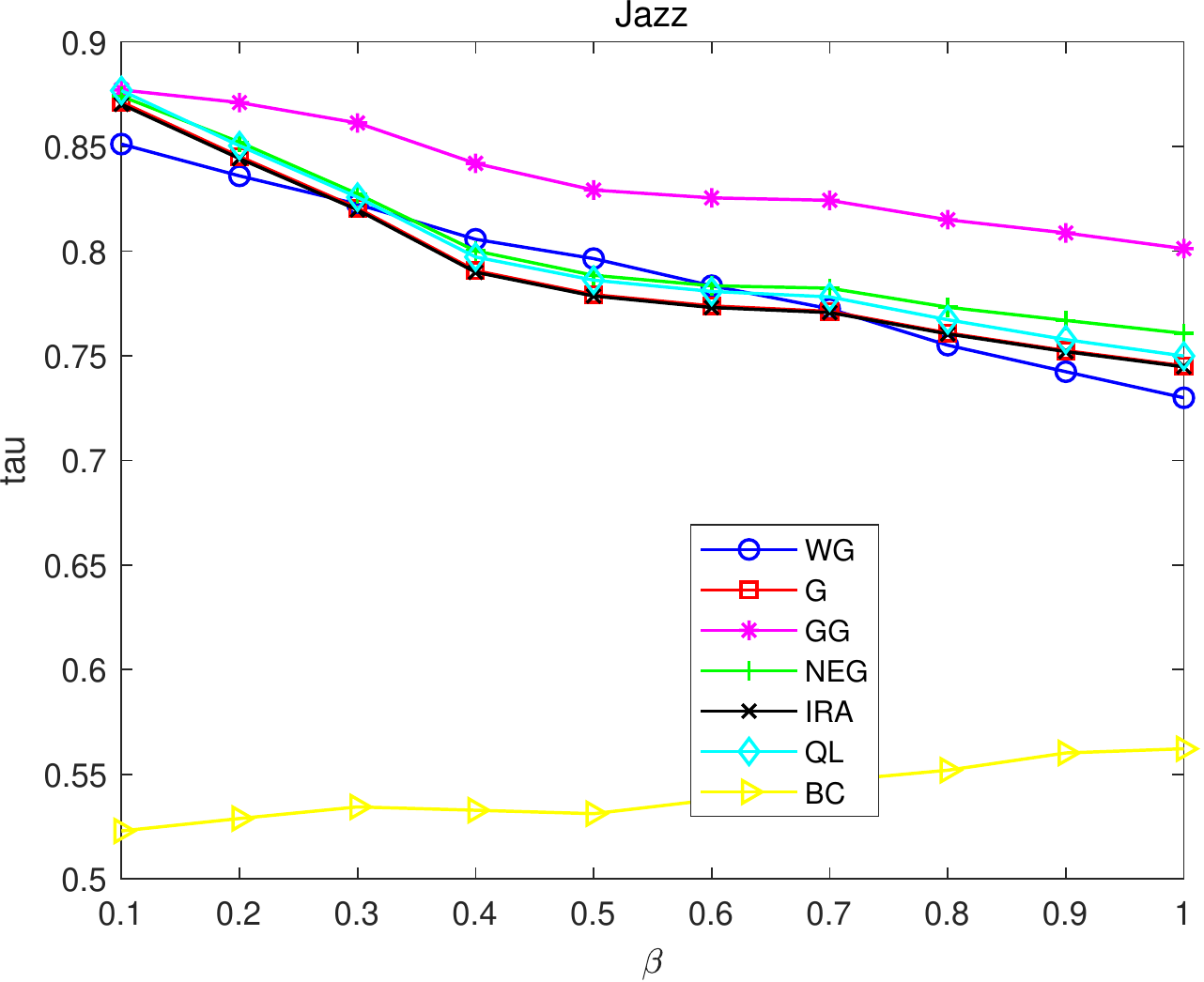}}
\subfigure{\includegraphics[scale=0.45]{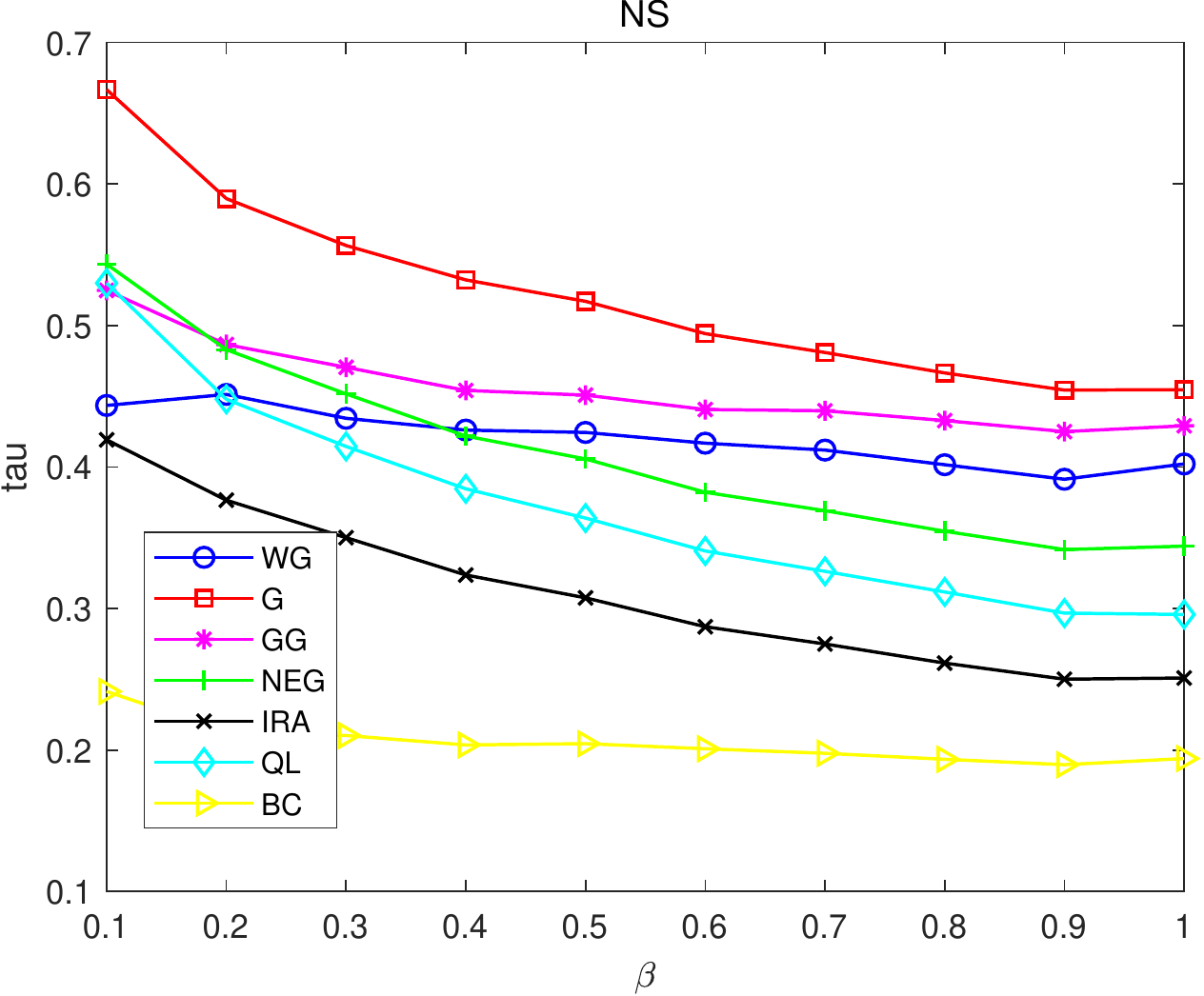}}
\subfigure{\includegraphics[scale=0.45]{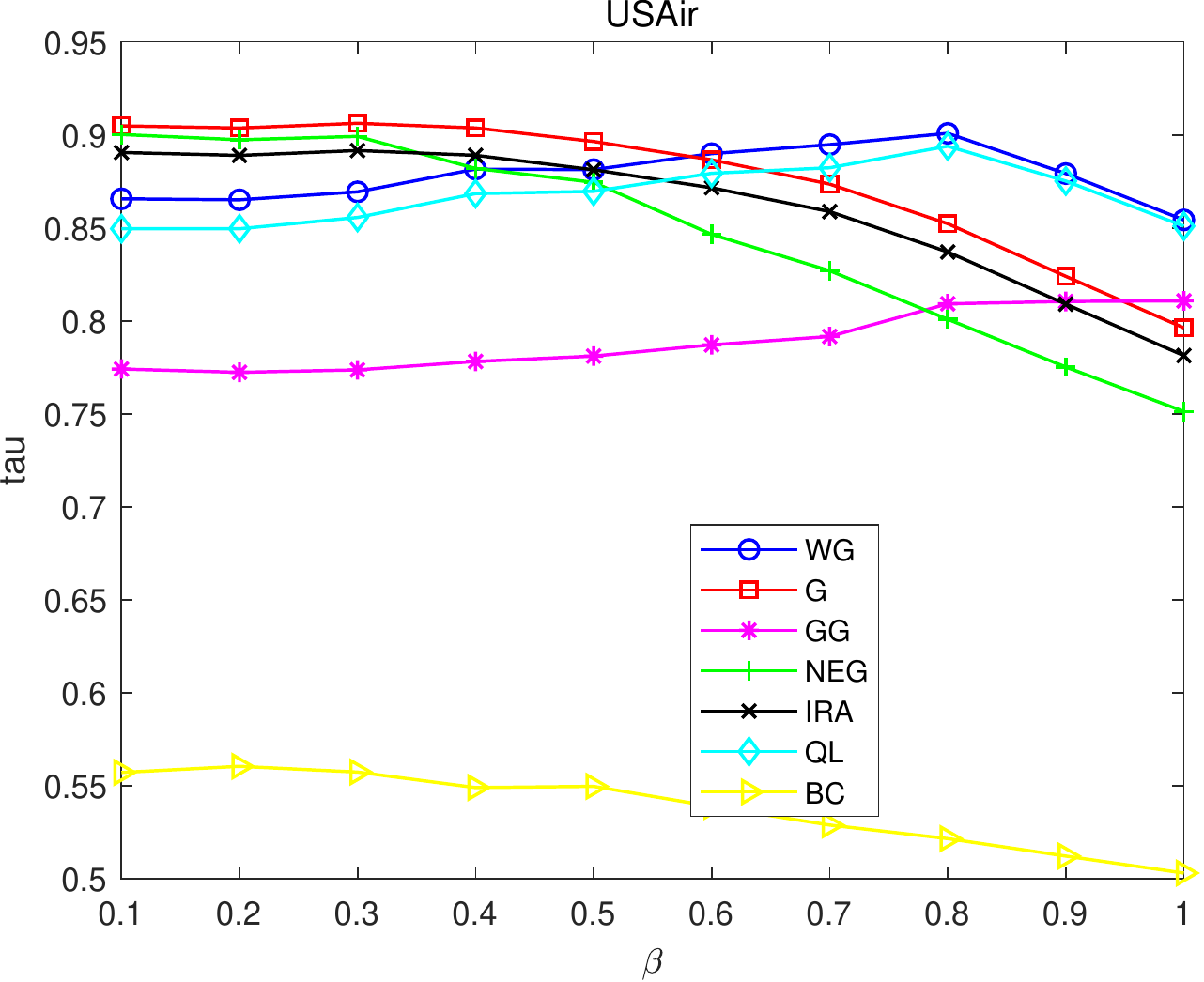}}
\subfigure{\includegraphics[scale=0.45]{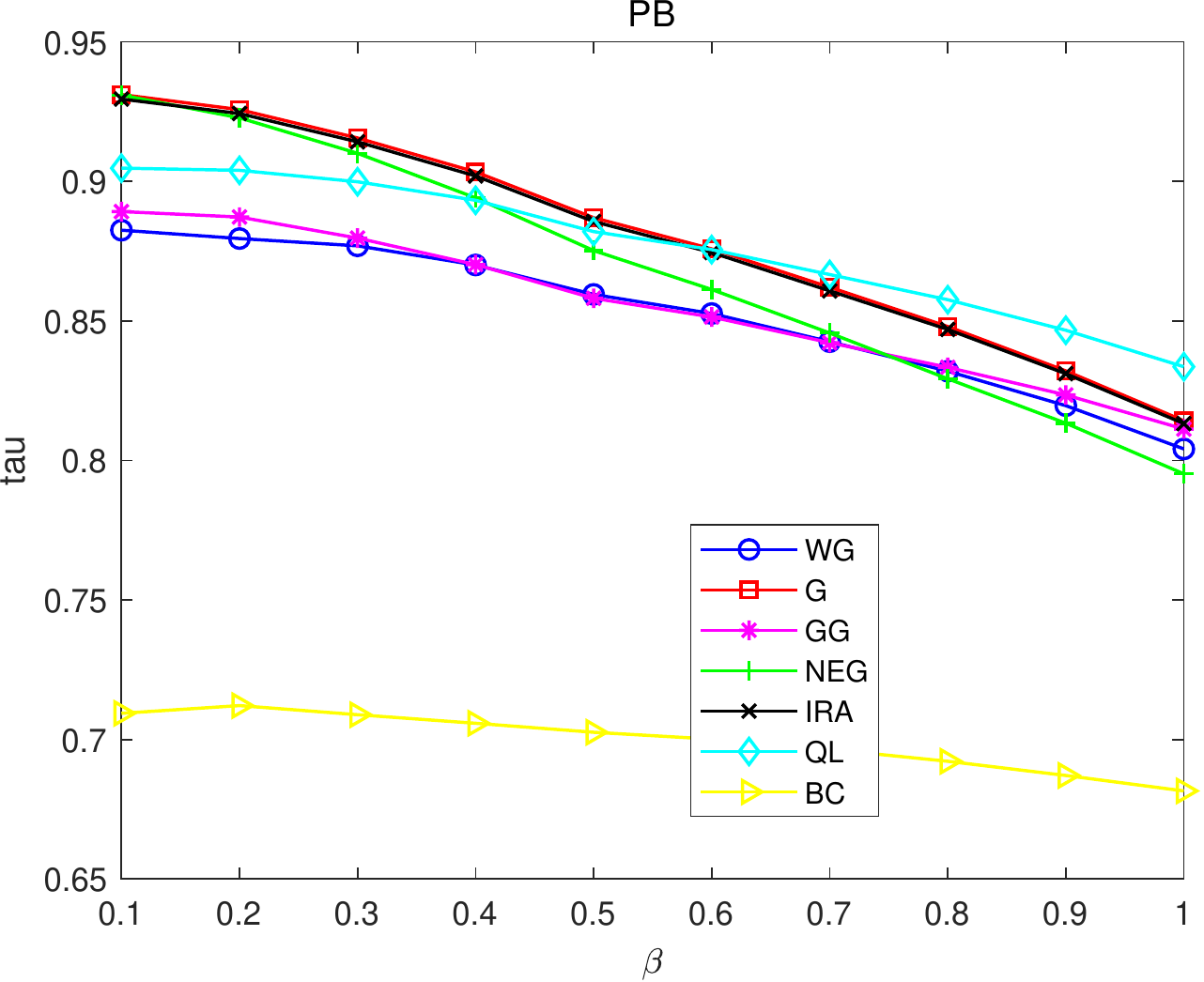}}
\subfigure{\includegraphics[scale=0.45]{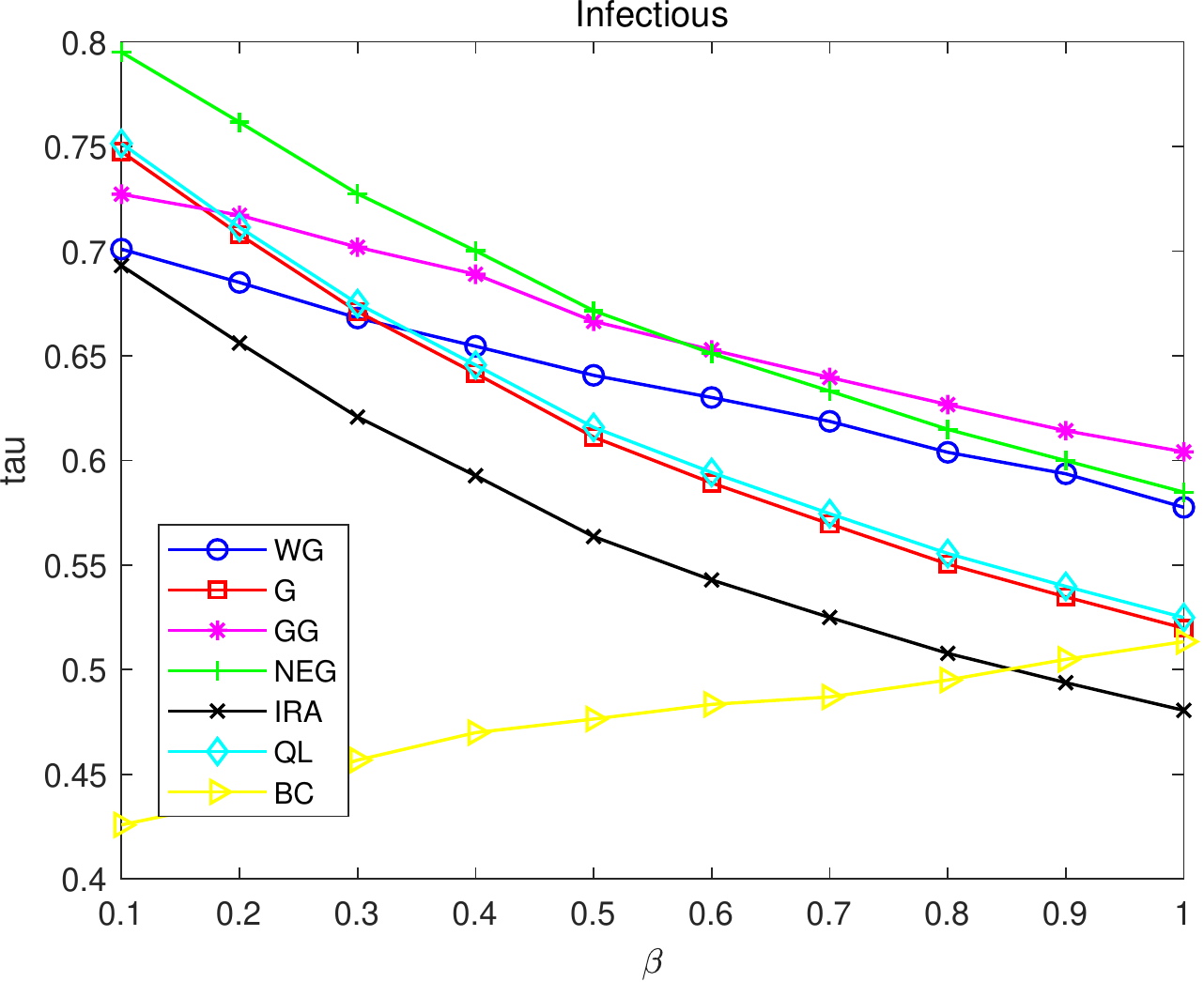}}
\subfigure{\includegraphics[scale=0.45]{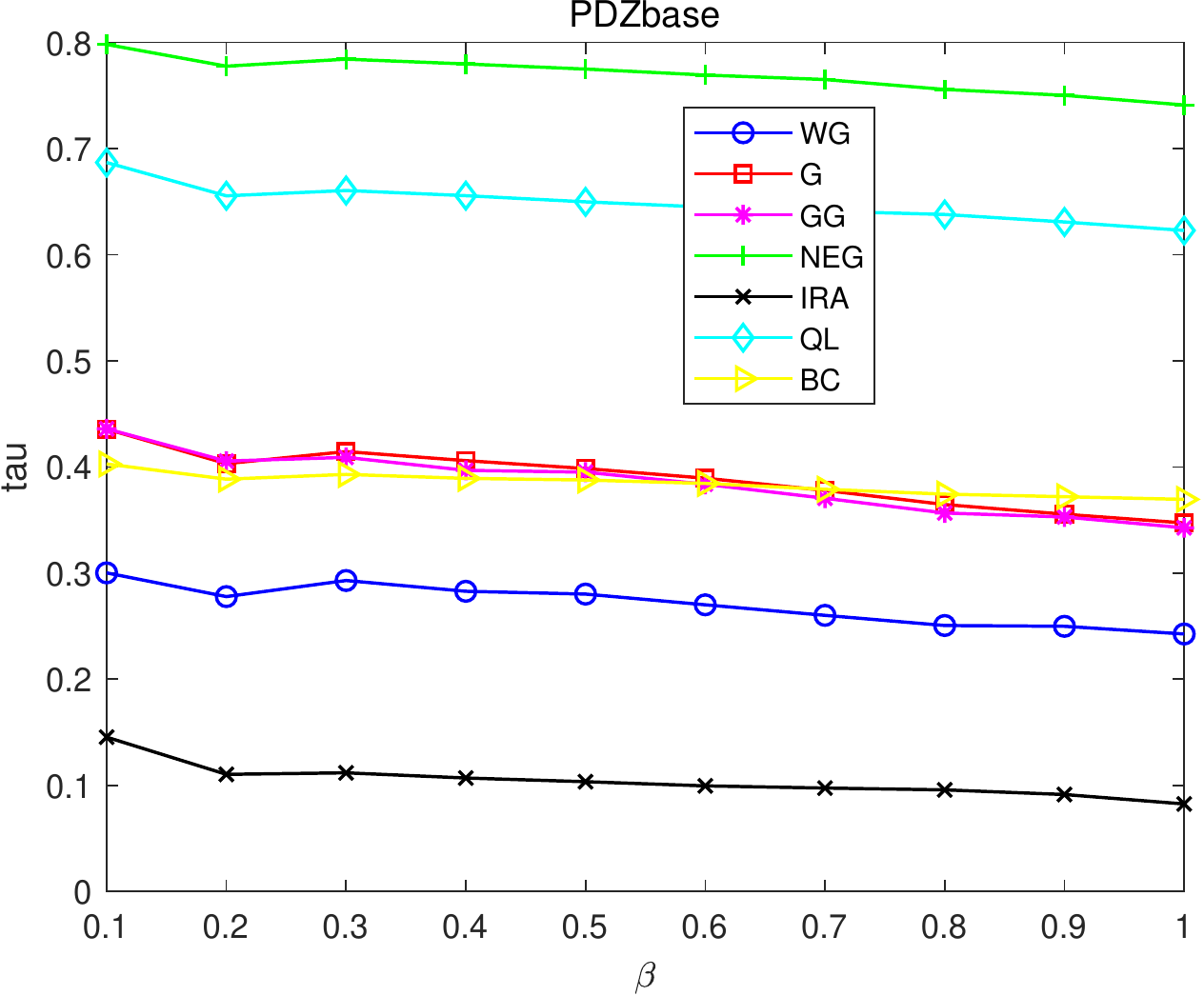}}
\subfigure{\includegraphics[scale=0.45]{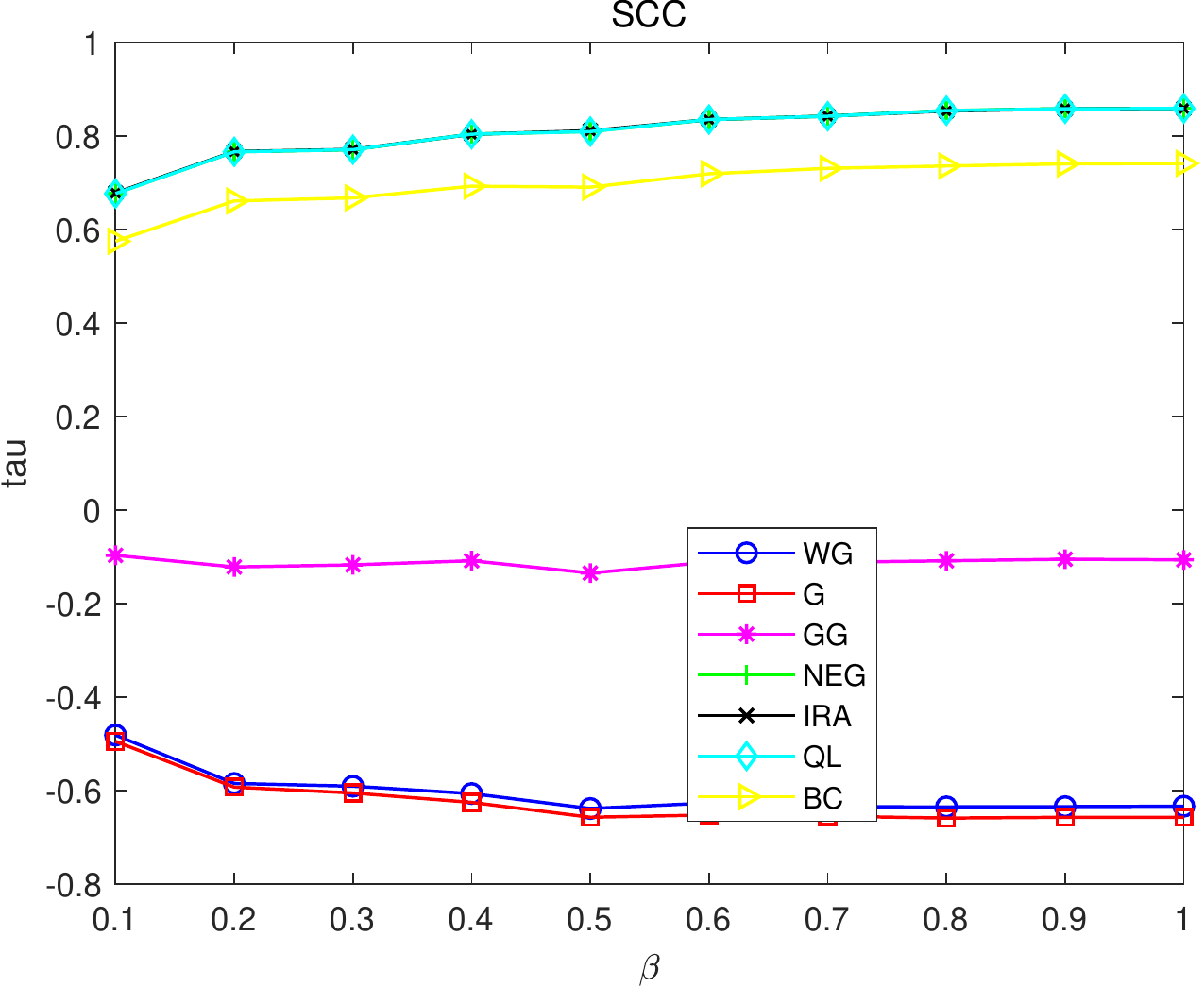}}
\subfigure{\includegraphics[scale=0.45]{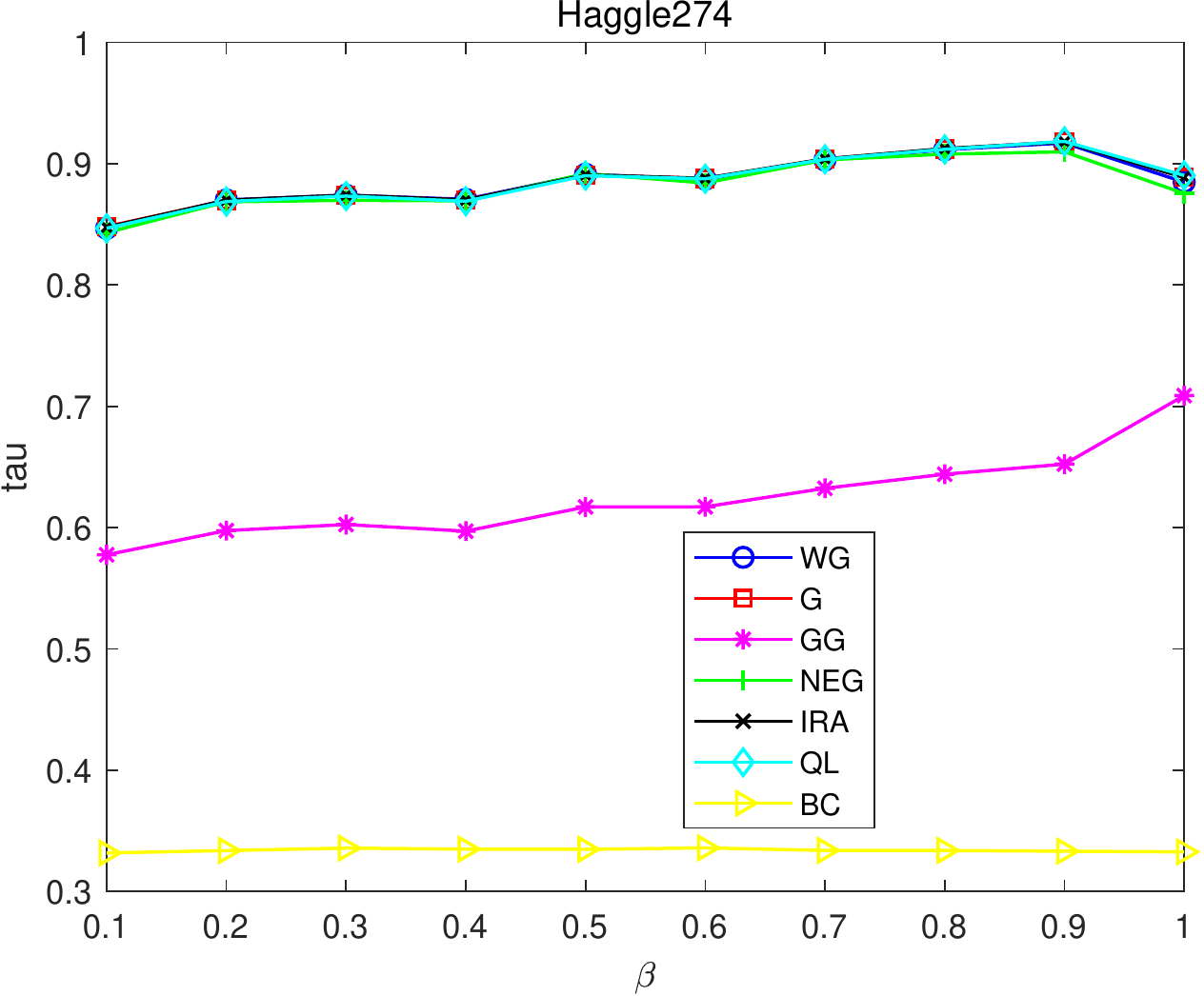}}
	\caption{Tau values of different centrality measures in networks}
\end{figure}

\subsection{The correlation with SI model}

The correlation between spreading ability of nodes in SI model and centrality can be modeled by Kendell correlation coefficient. If the correlation is high, the centrality is more efficient to find nodes that are influential in the network. The experiment are carried out 100 times independently and the $\beta$ varies from 0.1 to 1 and the step length is 0.1.

The results are shown in Figure 1. In Jazz network, the GG has the best performance while the IRA and G perform similarly. When $\beta<0.36$ and $\beta>0.6$, the tau of NEG is the second largest. In other cases, NEG is compatible with QL. In NS network, G performs the best while GG performs better than WG and NEG when $\beta>0.2$. As can be seen,  the tau of NEG is the second largest when $\beta<0.2$. In PB network, G and IRA perform best generally and NEG performs better than GG, WG and BC. In USAir network, our proposed method performs better than GG and BC and beaten by WG, G, IRA and QL. In infectious network, when $0.1 \leq \beta \leq 0.5$ the proposed outperforms other method. When $\beta \geq 0.6$, the proposed method is only outperformed by GG. In PDZbase network, NEG outperforms all other methods and is not even close. GG, G and BC are alomost the same. IRA has the worst performance. In SCC network, NEG performs similar with IRA and QL. BC is close to these three methods and GG performs better than G and WG. In Haggle274 network, our proposed method is comparable with IRA, WG, QL, G and has better performance than BC and GG. BC has the worst performance. As can be found, the tau value of BC is always lower than the others. In a nutshell, our proposed method performs compatible with most methods in most networks and performs best in some networks, which shows the feasibility and superiority of the proposed method. We can also conclude that GG has worse performance than G in 4 networks, better perfomance in 3 networks and comparable in 1 network. BC is the worst method among the six models.

\subsection{The infection ability of top-10 ranked nodes}
This experiment is executed 100 times independently. The SI model's simulation time $T$ equals to 25. The rate of infection $\beta$ is set as 0.1. Nodes which are ranked top-10 are elected as nodes are infected from the beginning. After $T$ steps, if there are more infected nodes, the centrality measure should be more effective. 

The results are shown in Figure 2. In Jazz network, when $3.6 < T < 3.8$, GG performs the best with NEG only performs worse than GG. G performs the worst among the four methods. In NS network, when $T < 16$, G performs best and WG performs worst. As can be seen, NEG performs the second best, which is slightly better than GG. In USAir network, the NEG performs the best while GG outperforms G and WG when $7.1 < t < 7.7$. When $t<10$, WG, GG, and NEG all perform better than G. In PB network, GG, NEG and G have similar performance and they outperforms WG. In infectious network, when $6.5 < T < 8.5$, GG performs slightly better than NEG and NEG beats G and WG by a large margin. In PDZbase network, NEG outperforms the GG and G by a large amount. GG is slightly better than G and WG has the worst performance. In SCC network, when $1.8 < T < 2.2$, NEG performs the best and GG performs better than G and WG. In Haggle274 network, four methods perform almost the same.

Generally speaking, NEG reaches the best performance in all networks, GG performs better than G and WG generally but is not as good as NEG. WG is beaten by all other methods. It shows the superiority of the proposed method. 

\begin{figure}
	\centering
	\subfigure{\includegraphics[scale=0.375]{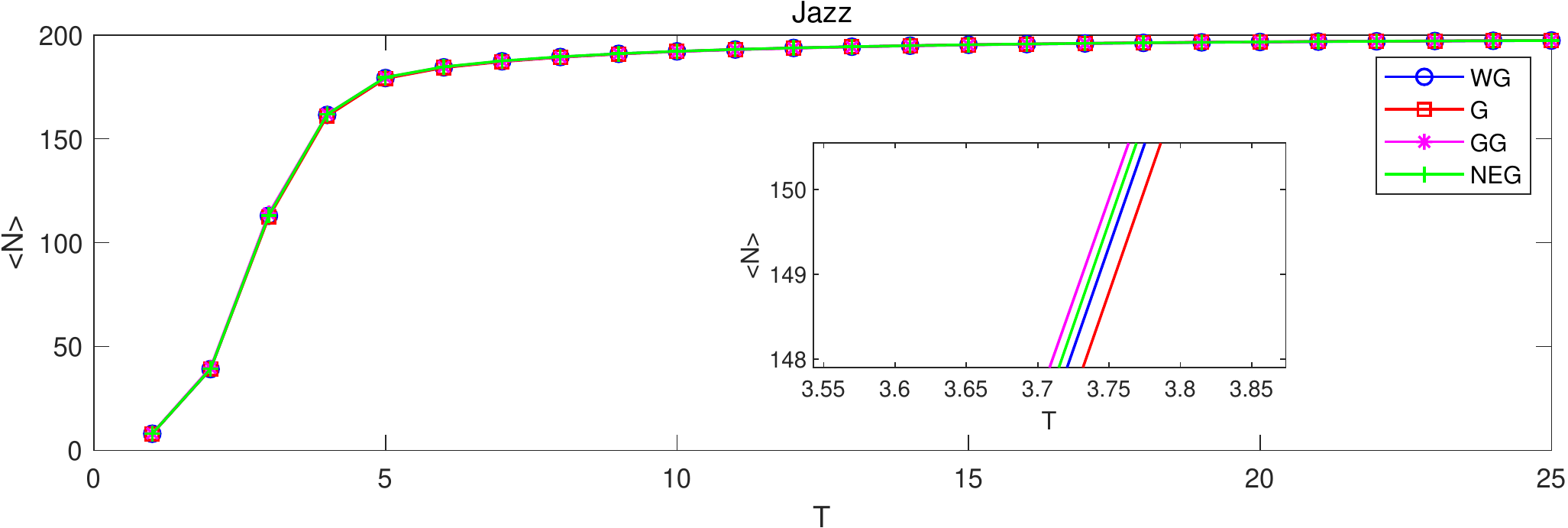}}
	\subfigure{\includegraphics[scale=0.375]{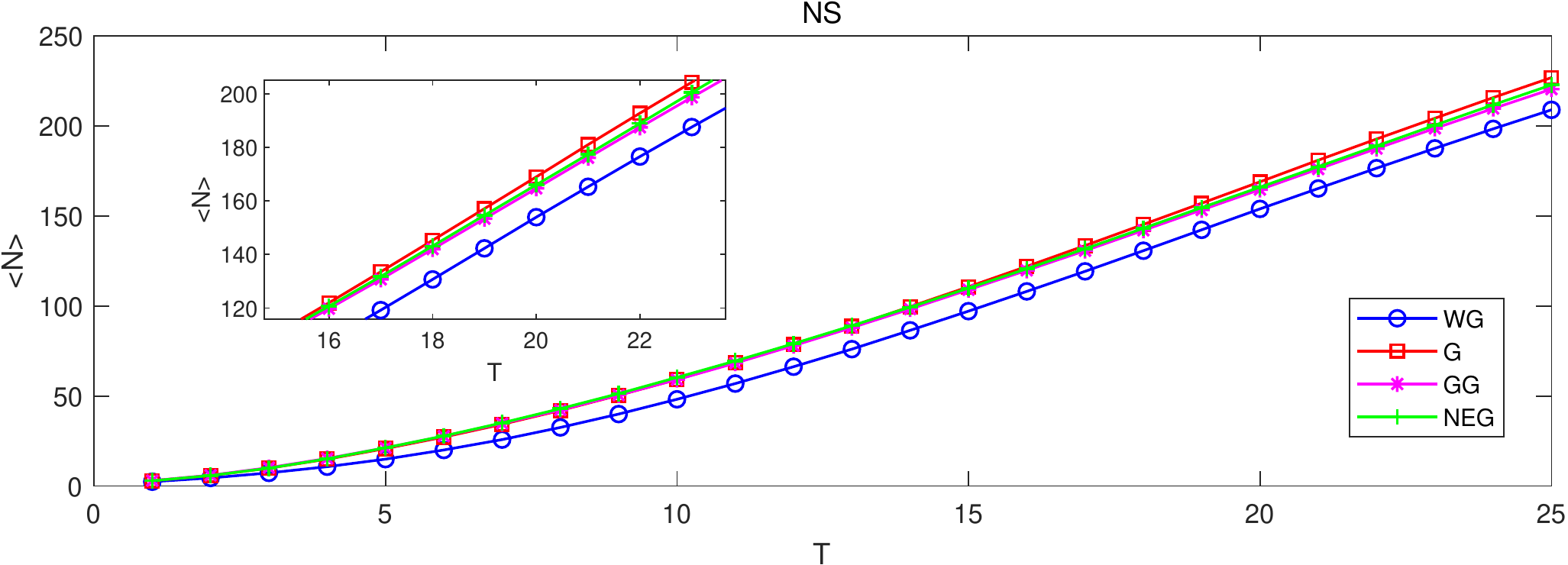}}
	\subfigure{\includegraphics[scale=0.375]{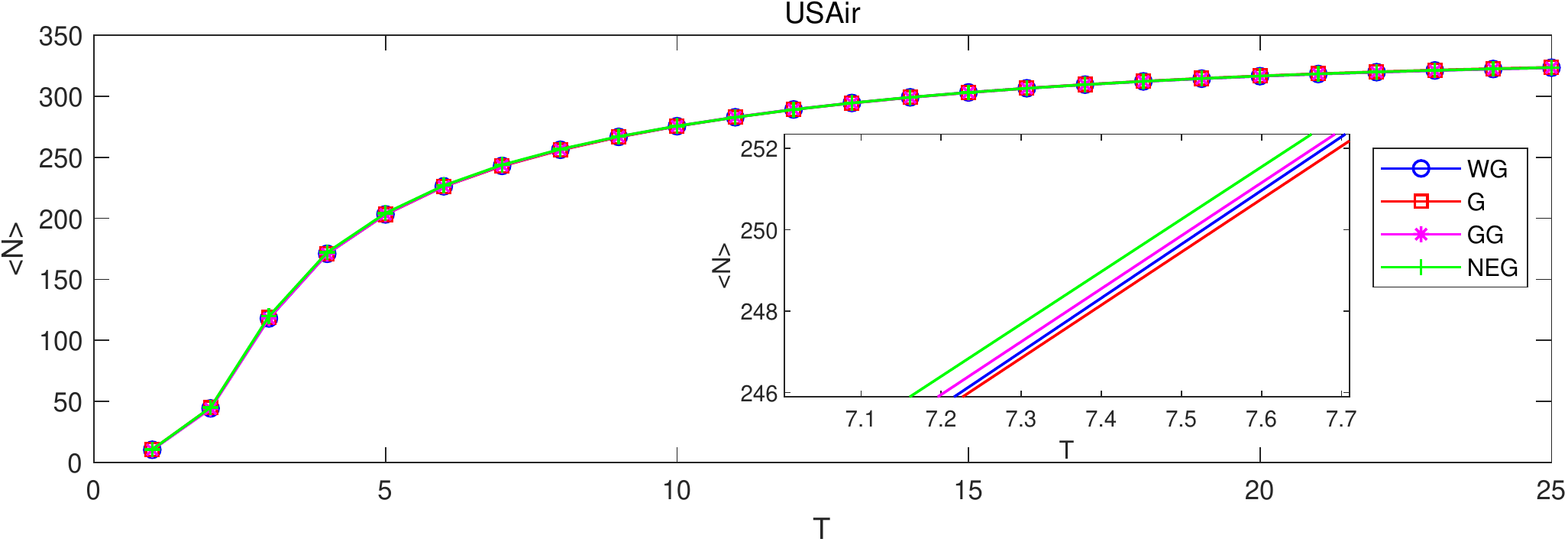}}
		\subfigure{\includegraphics[scale=0.375]{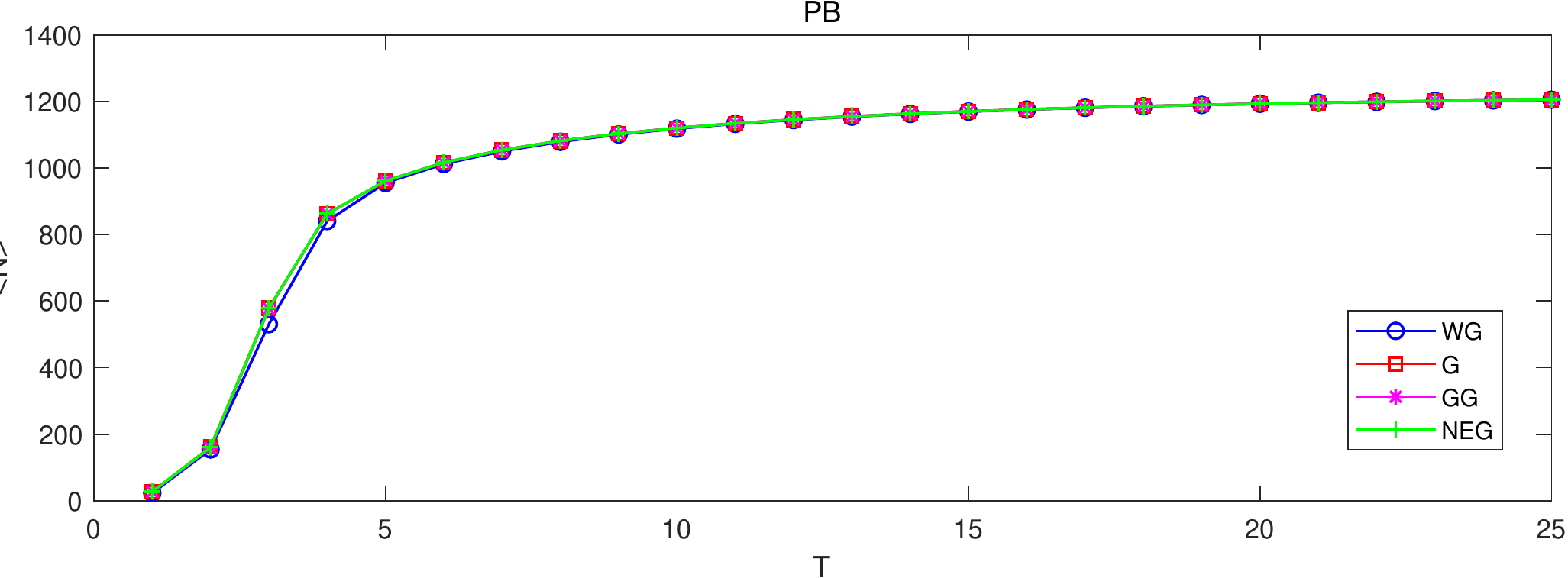}}
	\subfigure{\includegraphics[scale=0.375]{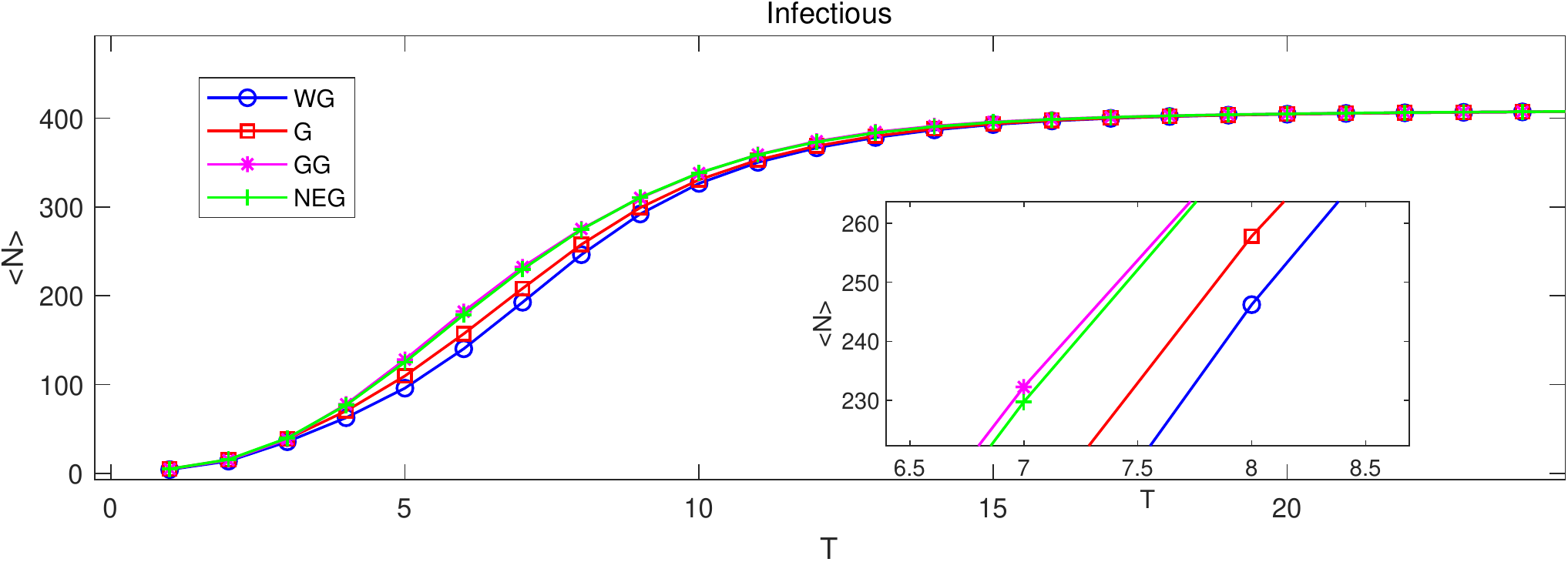}}
	\subfigure{\includegraphics[scale=0.375]{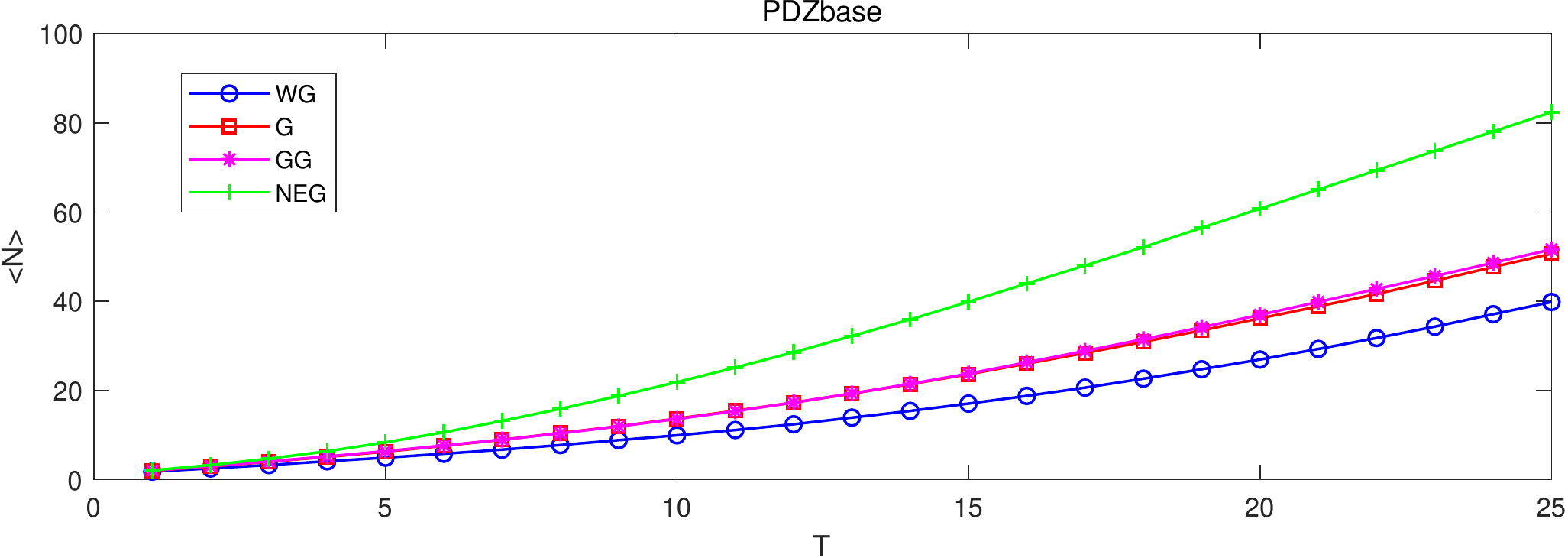}}
	\subfigure{\includegraphics[scale=0.375]{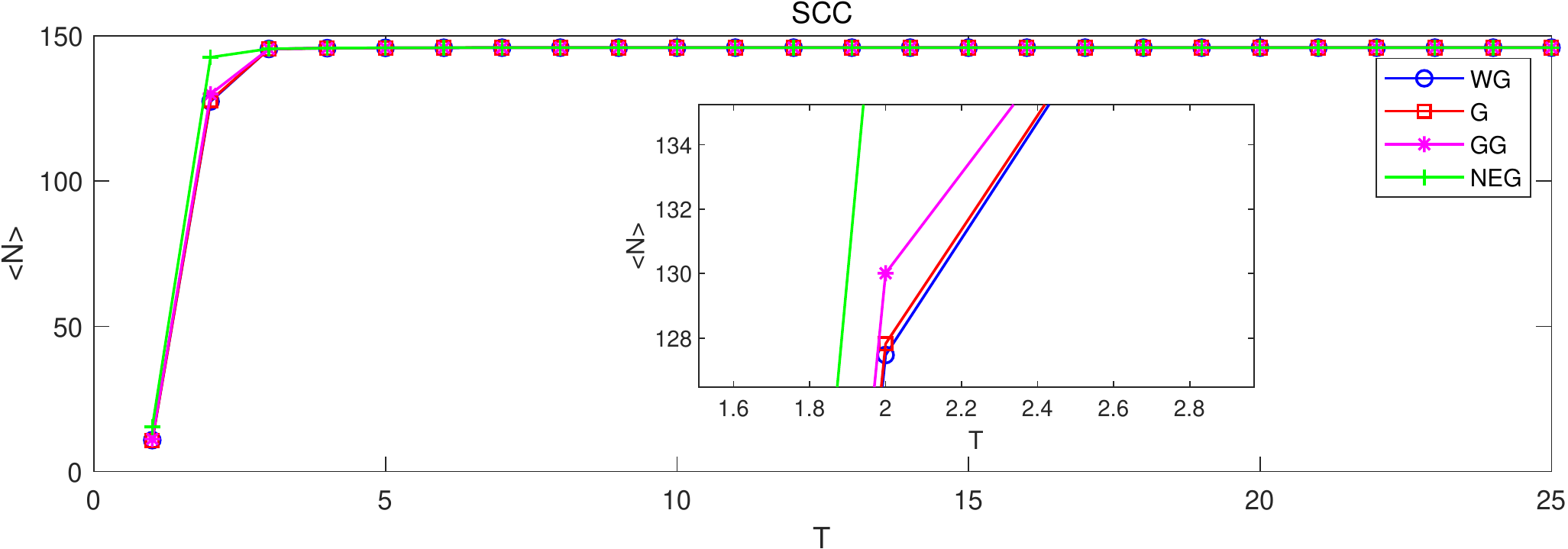}}
		\subfigure{\includegraphics[scale=0.35]{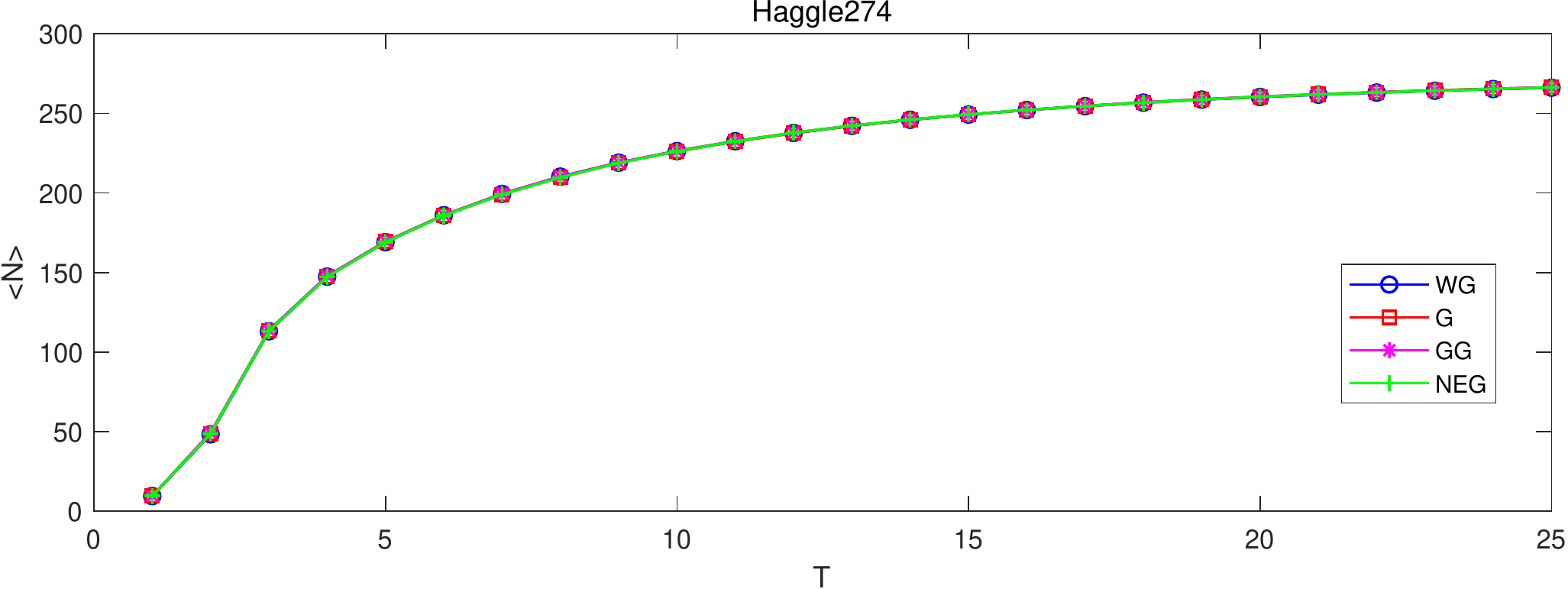}}
	\caption{Spreading abilities of different centrality measures in networks}
\end{figure}

\section{Conclusion}
A novel model for vital nodes identification named network efficiency gravity model is proposed in this paper, which integrates network efficiency and gravity model. The nodes' effect on the strucutral information on the whole network is taken into consideration in this model. The experiments on various networks come from the real-world prove the superiority of the proposed network efficiency gravity model. In the future, we will further study on how to decrease the time complexity of the proposed method to make the proposed method faster. Approximation algorithm of the proposed method may also be explored. The proposed algorithm may also be extended to other types of networks like multiplex networks, higher-order networks, temporal networks, bipartite networks and so forth. NEG can be used in many research directions like link prediction, community detection, graph clustering and so forth. Possible more practical applications with the proposed method is also worth being investigated. Due to the limits of the device, the proposed method can be tested by larger networks in the future if possible.

\section*{Acknowledgment}
%The authors greatly appreciate the reviews' suggestions and the editor's encouragement.
The work is partially supported by National Natural Science
Foundation of China (Grant No. 61973332), JSPS Invitational Fellowships for Research in Japan (Short-term).
\section*{References}
\bibliographystyle{elsarticle-num}
\bibliography{References}
\end{document}